# Concatenated Capacity-Achieving Polar Codes for Optical Quantum Channels

Laszlo Gyongyosi[*,a,b], Sandor Imre[a]

[a]*Quantum Technologies Laboratory, Department of Telecommunications*
*Budapest University of Technology and Economics*
2 Magyar tudosok krt, Budapest, H-1117, Hungary
[b]*Information Systems Research Group, Mathematics and Natural Sciences*
*Hungarian Academy of Sciences*
Budapest, H-1518, Hungary
[*]*gyongyosi@hit.bme.hu*

**Abstract:** We construct concatenated capacity-achieving quantum codes for noisy optical quantum channels. We demonstrate that the error-probability of capacity-achieving quantum polar encoding can be reduced by the proposed low-complexity concatenation scheme.

## 1. Introduction

Polar encoding is a revolutionary encoding scheme introduced by Arikan [1], and originally was developed for classical communication systems. Using polar codes, the symmetric capacity of any arbitrary classical communication channels can be provably achieved. The basic idea behind the construction of polar codes is channel selection, which is commonly called *polarization*: by assuming identical DMCs (*Discrete Memoryless Channel*), we can create two sets by means of an encoder [6]. "Good" channels are nearly noiseless, while "bad" channels have nearly zero capacity. Furthermore, for a large enough *n*, the fraction of good channels approaches the symmetric capacity of the original DMC. The polar encoding technique can also be extended to the quantum setting, and capacity-achieving quantum codes can be constructed [2], [4]. The main advantage of polar encoding is the low encoding and decoding complexities, $\mathcal{O}\left(n\log_2\left(n\right)\right)$, assuming block length *n*. The main drawback of polar encoding is that a long block length *n* is required to achieve low error probability; since, for a given *n*, the error probability of the scheme is $p_{err} = 2^{-\sqrt{n}}$. In this work, we show that, by using concatenated codes, the probability of error of the proposed capacity-achieving scheme can be reduced to $p_{err} = 2^{-n/\log_2(n)}$ in the block length *n*, essentially with the same computational complexity.

In the proposed code concatenation scheme, the polar codes are used as outer codes, while the inner codes are the quantum LDPC CSS (*Low-Density-Parity-Check Calderbank-Shor-Steane*) [2], [5] codes. Similar to polar codes, the LDPC coding scheme was also developed for classical systems, and is mainly used in wireless channels. The quantum LDPC code is the natural extension of classical LDPC codes in the quantum setting. The quantum LDPC codes also can be used to construct CSS codes, which play an important role in quantum error correction. In the proposed scheme we use quantum LDPC CSS codes. The quantum LDPC CSS codes are also capacity-achieving codes, however, a disadvantage of quantum LDPC CSS codes is that, for large block length, the "error floor" problem occurs [2]. Due to the "error floor" problem, in some regions the rate of the decay of the error probability will be lower and will converge with different speed to zero in the other regions, which could cause problems in the decoding process.

The proposed concatenation coding has two important advantages: First, the LDPC CSS codes decrease the error probability of the quantum polar codes. Second, by using quantum polar codes, the error-floor problem of quantum LDPC CSS codes can be completely eliminated. The proposed construction results in a capacity-achieving, high-performance, and low-error scheme, which can be applied in practical optical quantum communications.

An immediate practical application of the proposed concatenated code is to noisy optical quantum channels, quantum repeater networks or long-distance quantum communications, where capacity-achieving, low-complexity, and low-error-rate quantum codes are essential.

## 2. Proposed Quantum Code Concatenation Scheme

The proposed low-complexity capacity-achieving code concatenation scheme is illustrated in Fig. 1. The first block (outer code) is the quantum polar encoding scheme; the second (inner block) is the quantum LDPC CSS block.

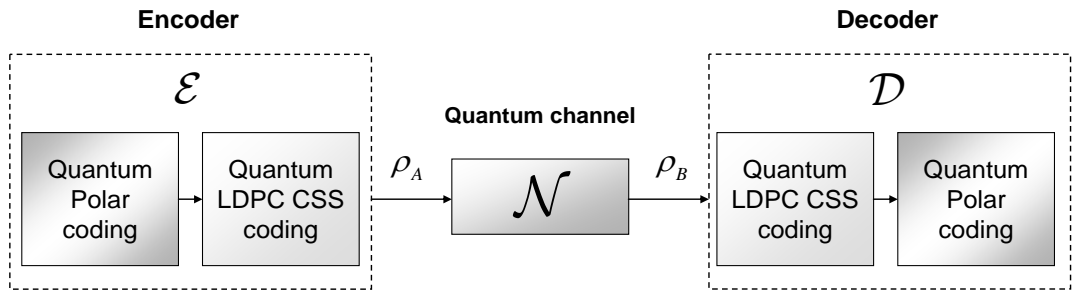

**Fig. 1.** The block diagram of the proposed code concatenation scheme. The first block is the quantum polar encoder; the second block creates quantum LDPC CSS quantum codes. Using the proposed scheme, the error probability of the polar encoding scheme can be decreased, and the error floor problem of the LDPC codes can be avoided. The concatenated code structure is a capacity-achieving code with low error probability.

The main advantage of the proposed code concatenation is that, while quantum polar codes can be used to achieve the capacity of the given noisy quantum channel, the quantum LDPC codes can help correct the block errors that occurred during the polar encoding. The error floor problem of quantum LDPC CSS codes can be overcome by the first block. The result of the proposed scheme is an improved capacity-achieving code with lower error probability and with no error floors.

## 3. Results

The main results of the error probabilities are shown in Fig. 2(a). The error probability of the polar encoding scheme is depicted by the dashed line (blue), while the error probability of the concatenated scheme is depicted by the solid (red) line. Significant improvement can be obtained in the error rate, since it is reduced to $p_{err} = 2^{-n/\log_2(n)}$, and the decay of the probability of error is almost exponential in the block length $n$, without the presence of the error floor problem. The computational complexity of the concatenated structure is nearly the same as the polar encoder block; it is changed from $\mathcal{O}(n \log_2(n))$ to $\mathcal{O}(n(\log_2(n))^2)$, which is still close to linear. In Fig. 2(b), the BER performances of the codes are compared, assuming an erasure quantum channel.

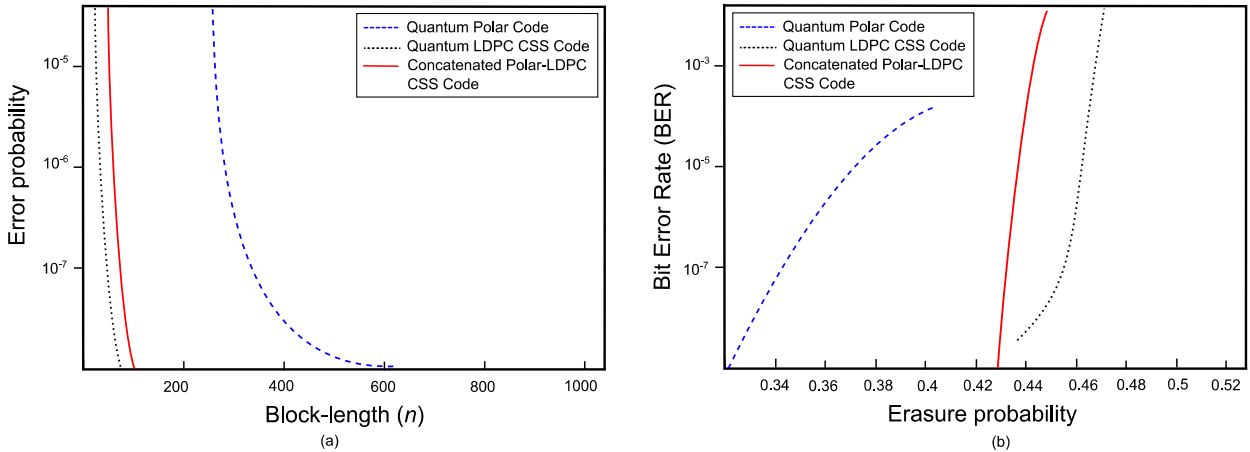

**Fig. 2**(a): The error probabilities of the quantum polar code, the quantum LDPC CSS code, and the concatenated code. Significant improvement achieved by the concatenation in comparison to the quantum polar code. 2(b): The BER rates assuming an erasure quantum channel. The BER performance of the concatenated code exceeds the performance of the polar code.

The performance of the proposed concatenation scheme exceeds the BER performance of the quantum polar code and has significantly lower error probabilities due to the quantum LDPC CSS codes. On the other hand, thanks to polar coding, the proposed concatenation scheme also avoids the error floor problem of LDPC codes, and the drawbacks of the two different encoding schemes can be completely eliminated in the concatenated structure.

## 4. Conclusions

In this work, we firstly introduced capacity-achieving concatenated quantum polar codes. As we showed, by using polar encoding, the main disadvantage of quantum LDPC CSS codes can be eliminated, and capacity-achieving codes can be constructed without the presence of the error floor problem. Moreover, the proposed concatenation scheme helps avoid the drawbacks of quantum polar encoding and quantum LDPC coding schemes and results in high-performance codes. In the proposed scheme, the channel polarization effect is exploited to construct codes that can achieve the capacity of arbitrary quantum channels with low encoding and decoding complexity. The quantum LDPC CSS codes were used to exponentially decrease the error probability of the polar encoding, which results in the use of significantly lower block lengths. Thanks to the proposed concatenation structure, the complexity remained in the same order and it is determined by only the quantum LDPC CSS block. The proposed concatenation scheme can be an efficient tool in noisy optical quantum communications, long-distance quantum communications, and in the development of quantum repeaters.

## 5. Acknowledgements

The results discussed above are supported by the grant TAMOP-4.2.2.B-10/1--2010-0009 and COST Action MP1006.